\documentclass[12pt]{article}
\usepackage{graphicx}
\usepackage{amsmath}
\usepackage{amssymb}
\usepackage{caption2}
\begin{document}
\bibliographystyle{prsty}
\begin{center}
{\large {\bf \sc{  Analysis of the $X(4350)$ as  a scalar $\bar{c}c$
and   ${D}_s^\ast {\bar {D}}_s^\ast$ mixing  state with  QCD sum rules  }}} \\[2mm]
Zhi-Gang Wang \footnote{E-mail:wangzgyiti@yahoo.com.cn.  }    \\
 Department of Physics, North China Electric Power University,
Baoding 071003, P. R. China
\end{center}

\begin{abstract}
In this article, we assume that the narrow structure $X(4350)$ is a
scalar $\bar{c}c-{D}_s^\ast {\bar {D}}_s^\ast$ mixing state, and
study its mass using the QCD sum rules.  The numerical result
$M_X=(4.37\pm0.15)\,\rm{GeV}$ is in good agreement with the
experimental data, the $X(4350)$ may be a scalar
$\bar{c}c-{D}_s^\ast {\bar {D}}_s^\ast$ mixing state. Other
possibility, such as a scalar (tensor) $cs\bar{c}\bar{s}$ tetraquark
state is not excluded.
\end{abstract}

 PACS number: 12.39.Mk, 12.38.Lg

Key words: X(4350), QCD sum rules

\section{Introduction}
In 2009, the CDF collaboration   observed   a narrow structure
(which is denoted as the $Y(4140)$  now) near the $J/\psi\phi$
threshold with statistical significance in excess of 3.8 standard
deviations in exclusive decays $B^+\to J/\psi\phi K^+$  produced in
$\bar{p} p $ collisions at $\sqrt{s}=1.96 \,\rm{TeV}$
\cite{CDF0903}. The mass and  width  are $(4143.0\pm2.9\pm1.2)\,\rm{
MeV}$ and $(11.7^{+8.3}_{-5.0}\pm3.7)\, \rm{MeV}$ respectively. The
narrow structure  $Y(4140)$ is very similar to the charmonium-like
state $Y(3930)$ near the $J/\psi \omega$ threshold
\cite{Belle2005,Babar2008}. The mass and width of the $Y(3930)$ are
$(3914.6^{+3.8}_{-3.4}\pm 2.0) \,\rm{MeV}$ and $(34^{+12}_{-8}\pm
5)\,\rm{MeV}$  respectively \cite{Babar2008}.

There have been several explanations for the nature of the   narrow
structure $Y(4140)$, such as a  $D_s^\ast {\bar D}_s^\ast$ molecular
state \cite{LiuMolecule,Mahajan0903,BranzMolecule,NielsenMolecule,
DingMolecule,HuangMolecule,OsetMolecule}, an exotic
($J^{PC}=1^{-+}$) hybrid charmonium \cite{Mahajan0903}, a $c \bar c
s \bar s$ tetraquark state \cite{Stancu4},
  the effect of the  $J/\psi \phi$ threshold  \cite{NoResonance},
 or none a conventional
charmonium state \cite{LiuRescatter}  nor a scalar $D_{s}^{\ast}
\bar{D}_{s}^{\ast}$ molecular state \cite{Wang0903,Wang0907}, etc.
Assuming the $Y(4140)$ is a $D_{s}^{\ast} \bar{D}_{s}^{\ast}$
molecular state with $J^{PC}=0^{++}$ or $2^{++}$, Branz et al
 predict its two-photon decay width is of
order $1\,\rm{ KeV}$  \cite{BranzMolecule}.

Recently,  the Belle collaboration measured the process $\gamma
\gamma \to \phi J/\psi$ for the $\phi J/\psi$ invariant mass
distributions between the threshold and $5\,\rm{GeV}$ based on a
data sample of $825\,\rm{fb}^{-1}$, and observed a narrow peak of
$8.8^{+4.2}_{-3.2}$  events with a significance of $3.2$ standard
deviations \cite{Belle4350}. The mass and width of the structure
(denoted as $X(4350)$) are $(4350.6^{+4.6}_{-5.1}\pm 0.7)\,\rm{MeV}$
and $(13.3^{+17.9}_{-9.1}\pm 4.1)\,\rm{MeV}$ respectively. No signal
for the $Y(4140)\to \phi J/\psi$ structure was observed, this
disfavors the scenario of the $Y(4140)$ as a $D_{s}^{\ast}
\bar{D}_{s}^{\ast}$ molecular state.

The possible quantum numbers for a state $X$ decaying into $J/\psi
\phi$ are $J^{PC}=0^{-+},0^{++},1^{-+},2^{++}$, the corresponding
strong  interactions can be described by the following
phenomenological Lagrangian,
\begin{eqnarray}
\mathcal{L}_{0^{-+}}&=& g \epsilon^{\mu\nu\alpha\beta}\partial_\mu
\psi_\nu \partial_\alpha \phi_\beta X \, ,  \nonumber \\
\mathcal{L}_{1^{-+}}&=& g \epsilon^{\mu\nu\alpha\beta}
(\psi_\mu \partial_\nu\phi_\alpha-\phi_\mu \partial_\nu \psi_\alpha)  X_\beta \, ,  \nonumber \\
\mathcal{L}_{0^{++}}&=& g
\psi_\mu  \phi^\mu X \, ,  \nonumber \\
\mathcal{L}_{2^{++}}&=& g X^{\mu\nu} \psi_\mu \phi_\nu \, ,
\end{eqnarray}
 the strong coupling constants $g$ can be fitted
phenomenologically or calculated by some theoretical  approaches,
for example, the QCD sum rules.

In Ref.\cite{HuangMolecule}, Zhang and Huang study the
$Q\bar{s}\bar{Q}s$ and $Q\bar{s}\bar{Q}'s$ molecular states in a
systematic way using the QCD sum rules  before the Belle
experiment, the mass of the $D_s^{*+}\bar{D}_{s0}^{-}$ molecular
state is $(4.36 \pm 0.08)\,\rm{GeV}$, which is consistent with
experimental data \cite{Belle4350}. Such a state has $J^P=1^-$ and
no definite charge conjugation. In Ref.\cite{Albuquerque1001},
Albuquerque et al
 re-study the exotic $D_s^{*+}\bar{D}_{s0}^{-}-\bar{D}_s^{*-}D_{s0}^{+}$ molecular state
with $J^{PC}=1^{-+}$ by taking into account the contributions from
the vacuum condensates up to dimension-8, the prediction
$M_{D_s^*{D}_{s0}}=(5.05\pm 0.19)\,\rm{GeV}$    disfavors the
 scenario of the $X(4350)$  as a  $D_s^*\bar{D}_{s0}$ molecular state.

In Ref.\cite{Drenska0902}, Drenska et al study the exotic tetraquark
states of the kind $cs\bar{c} \bar{s}$ by computing their spectrum
and decay modes within a constituent diquark-antidiquark model, the
predictions $M_{0^{-+}}=4277,4312\,\rm{MeV}$ and
$M_{1^{-+}}=4321,4356\,\rm{MeV}$ are consistent with the
experimental data \cite{Belle4350}. On the other hand, the flux-tube
model \cite{FluxTube-1,FluxTube-2} and the Lattice QCD
\cite{Latt-1,Latt-2,Latt-3} predict that the masses of the low lying
hybrid charmonium states are about $(4.0-4.2)\,\rm{ GeV }$ and
$(4.0-4.4) \, \rm{GeV}$ respectively, which are also consistent with
the experimental data \cite{Belle4350}. However, the decay of a
hybrid to two photons is generically forbidden \cite{Page1997}.

In Ref.\cite{Stancu4}, Stancu study the mass spectrum of the $c
\bar{c} s \bar{s}$ tetraquarks using a simple quark model with
chromo-magnetic interaction and observe that the  $Y(4140)$  may be
the strange partner of the $X(3872)$,  the prediction for the mass
of the $2^{++}$ tetraquark state is consistent with the $X(4350)$.
As noticed  by the author, the amplitude of the singlet-singlet
component seems too large comparing with the octet-octet component.

 In
Ref.\cite{Liu0911}, Liu et al discuss   the possibility that the
$X(4350)$ is an excited $P$-wave charmonium state $\chi_{c2}''$  by
studying  the strong decays of the $P$-wave charmonium states with
the $^3P_0$ model.

The CDF and Belle collaborations analyze the experimental data by
assuming  the vector mesons $J/\psi$ and $\phi$  have a relative
$S$-wave \cite{CDF0903,Belle4350}, so we will not  focus on  the
scenarios of the $X(4350)$ as the $0^{-+}$ and $1^{-+}$ tetraquark
state or hybrid charmonium.

\begin{table}
\begin{center}
\begin{tabular}{|c|c|c|c|c|c|c|}
\hline\hline
 Nature  & $J^{PC}$ &  $M_X(\rm{GeV})$&Decay Channels&RW& References \\ \hline
        $D^{*}_s \bar{D}_{s0}$  &$1^{-?}$ &$4.36 \pm 0.08$& $J/\psi\phi$&$P$ &\cite{HuangMolecule}\\ \hline
      $D^{*}_s \bar{D}_{s0}-\bar{D}^{*}_s D_{s0}$  &$1^{-+}$ &$5.05 \pm 0.19$& $J/\psi \phi$&$P$ &\cite{Albuquerque1001}\\ \hline
     $cs\bar{c}\bar{s}$  &$1^{-+}$ &$4.321/4.356$& $J/\psi \phi$&$P$&\cite{Drenska0902}\\ \hline
       $cs\bar{c}\bar{s}$  &$2^{++}$ &$4.343/4.359$& $J/\psi \phi;\,D_s^*\bar{D}_s^*;\,D_s\bar{D}_s$&$S;D$&\cite{Stancu4}\\ \hline
      $\chi''_{c2}$  &$2^{++}$ && $D\bar{D};\,D\bar{D}^*;\,D^*\bar{D}^*;\cdots$&$S;P;D$&\cite{Liu0911}\\ \hline
    $cs\bar{c}\bar{s}$  &$0^{-+}$ &$4.277/4.312$& $J/\psi\phi;\,D_s^*\bar{D}_s^*$&$P$&\cite{Drenska0902}\\ \hline
    $cs\bar{c}\bar{s}$  &$0^{++}$ &$4.45\pm0.16$& $J/\psi\phi;\,D_s^*\bar{D}_s^*$&$S$&\cite{WangScalar41,WangScalar42}\\ \hline
    $c\bar{c}-D_s^*\bar{D}_s^*$  &$0^{++}$ &$4.37\pm0.15$& $J/\psi \phi;\,D_s\bar{D}_s;\,D_s^*\bar{D}_s^*$&$S$&This Work\\ \hline
    \hline
\end{tabular}
\end{center}
\caption{ The possible explanations for the nature of the $X(4350)$,
where the RW stands for the relative wave of the final state
mesons.}
\end{table}

 In Refs.\cite{WangScalar41,WangScalar42},  we study the mass spectrum of the scalar hidden charm and
hidden bottom tetraquark states which consist of the scalar-scalar
type,  axial-vector-axial-vector  type and  vector-vector type
diquark pairs in a systematic way using the QCD sum rules, the
scalar-scalar type and axial-vector-axial-vector  type scalar
$c\bar{c}s\bar{s}$ tetraquark states have masses about
$(4.45\pm0.16)\,\rm{GeV}$, the lower bound of the masses is
consistent with the $X(4350)$, we cannot exclude that the $X(4350)$
is a scalar $c\bar{c}s\bar{s}$ tetraquark state.
 In Refs.\cite{Wang0903,Wang0907}, we study the ${D}^\ast {\bar {D}}^\ast$, ${D}_s^\ast
{\bar {D}}_s^\ast$, ${B}^\ast {\bar {B}}^\ast$ and ${B}_s^\ast {\bar
{B}}_s^\ast$ molecular states in a systematic way  using  the QCD
sum rules. The numerical result  is  inconsistent with the
experimental data, the $D_s^\ast {\bar D}_s^\ast$  is probably a
virtual state and not related to the meson $Y(4140)$. In this
article, we study the $X(4350)$ as a linear superposition of a
scalar charmonium state $c\bar{c}$ and a virtual state $D_s^\ast
{\bar D}_s^\ast$ using the QCD sum rules  \cite{SVZ79,Reinders85}.
In Table 1, we present  the possible explanations for the nature of
the $X(4350)$.

The article is arranged as follows:  we derive the QCD sum rules for
the narrow structure $X(4350)$ in   Sect.2;
 in Sect.3, we present the numerical results and discussions; and Sect.4 is reserved for our
conclusions.

\section{QCD sum rules for  the $X(4350)$ as a mixing state }
 In the following, we write down  the
two-point correlation function $\Pi(p)$  in the QCD sum rules,
\begin{eqnarray}
\Pi(p)&=&i\int d^4x e^{ip \cdot x} \langle
0|T\left\{J(x)J^{\dagger}(0)\right\}|0\rangle \, , \\
J(x)&=& \frac{J_{1}(x)+J_2(x)}{\sqrt{2}} \, , \nonumber\\
J_1(x)&=&\bar{c}(x)\gamma_\mu s(x) \bar{s}(x)\gamma^\mu c(x) \, , \nonumber \\
J_2(x)&=&-\frac{\langle\bar{s}s\rangle}{3}\bar{c}(x)c(x) \, ,
\end{eqnarray}
where the  $J_2(x)$ is the normalized two-quark current
\cite{Sugiyama2007}.

We can insert  a complete set of intermediate hadronic states with
the same quantum numbers as the current operator $J(x)$  into the
correlation function $\Pi(p)$  to obtain the hadronic representation
\cite{SVZ79,Reinders85}. After isolating the ground state
contribution from the pole term of the lowest state $X$, we get the
following result,
\begin{eqnarray}
\Pi(p)&=&\frac{\lambda_{X}^2}{M_{X}^2-p^2} +\cdots \, \, ,
\end{eqnarray}
where the pole residue (or coupling) $\lambda_X$ is defined by
\begin{eqnarray}
\lambda_{X} &=& \langle 0|J(0)|X(p)\rangle \, .
\end{eqnarray}

The two-quark current $\bar{c}(x) c(x)$ has non-vanishing coupling
with the charmonia $\chi_{c0}(1P)$, $\chi_{c0}(2P)$,
$\chi_{c0}(3P)$, $\cdots$; while the molecule type  current
$\bar{c}(x)\gamma_\mu s(x) \bar{s}(x)\gamma^\mu c(x)$ has
non-vanishing coupling with the molecular states  ${D}_s^\ast {\bar
{D}}_s^\ast$, ${D}_s^\ast {\bar {D'}}_s^\ast$, ${D'}_s^\ast {\bar
{D'}}_s^\ast$, $\cdots$ and the scattering states ${D}_s^\ast-{\bar
{D}}_s^\ast$, ${D}_s^\ast -{\bar {D'}}_s^\ast$, ${D'}_s^\ast- {\bar
{D'}}_s^\ast$, $\cdots$ \cite{PDG}. We cannot distinguish those
contributions and study them exclusively.  In this article, we take
the assumption that the interpolating current $J(x)$ couples to a
particular resonance, which is a special  superposition of the
scalar charmonia $\chi_{c0}(1P)$, $\cdots$ and the virtual molecular
states ${D}_s^\ast {\bar {D}}_s^\ast$, $\cdots$. In other words, we
take a single pole approximation, the pole embodies  the net
effects.

 We carry out the operator product
expansion for the correlation function $\Pi(p)$    at  the large
space-like momentum region $p^2\ll 0$,
\begin{eqnarray}
\Pi(p)&=& \frac{1}{2}\Pi_{11}(p)+\frac{\langle
\bar{s}s\rangle^2}{6}\Pi_{22}(p) \, ,
\end{eqnarray}
where
\begin{eqnarray}
\Pi_{11}(p)&=& i\int d^4x e^{ip \cdot x} \langle
0|T\left\{J_1(x)J_1^{\dagger}(0)\right\}|0\rangle=
\int_{\Delta^2}^{s_0} ds
\frac{\rho_{11}(s)}{s-p^2}+\cdots \, ,\nonumber \\
\Pi_{22}(p)&=&i\int d^4x e^{ip \cdot x} \langle
0|T\left\{J_2(x)J_2^{\dagger}(0)\right\}|0\rangle=
\int_{\Delta^2}^{s_0} ds
\frac{\rho_{22}(s)}{s-p^2}+\cdots \, ,\nonumber \\
\rho_{22}(s)&=&\frac{9}{4\pi^2}\int_{x_i}^{x_f}dx
x(1-x)(s-\widetilde{m}_c^2)\nonumber
\\
&&+\frac{1}{8}\langle\frac{\alpha_sGG}{\pi}\rangle
\int_0^1dx\left[1-\frac{(x^2-x+1)\widetilde{m}_c^2}{x(1-x)M^2}
\right]\delta(s-\widetilde{m}_c^2) \, ,
\end{eqnarray}
the explicit expression of the spectral density $\rho_{11}(s)$ can
be found in Refs.\cite{Wang0903,Wang0907}, $\Delta^2=4(m_c+m_s)^2$,
$\widetilde{m}_c^2=\frac{m_c^2}{x(1-x)}$,
$x_{f}=\left(1+\sqrt{1-\frac{4m_c^2}{s}}\right)/2$,
$x_{i}=\left(1-\sqrt{1-\frac{4m_c^2}{s}}\right)/2$. In calculation,
we use the  Fierz re-ordering in the color space  and Dirac spin
space to express the correlation functions $\Pi_{12}(p)$ and
$\Pi_{21}(p)$ in terms of the $\Pi_{22}(p)$. In this article, we
carry out the operator product expansion to the vacuum condensates
adding up to dimension-10 and
 take the assumption of vacuum saturation for the  high
dimensional  vacuum condensates, they  are always
 factorized to lower condensates with vacuum saturation in the QCD sum
 rules, and   factorization works well in  large $N_c$ limit.

Once analytical result is obtained,   then we can take the
quark-hadron duality and perform the Borel transform  with respect
to the variable $P^2=-p^2$, finally we obtain  the following  sum
rule:
\begin{eqnarray}
\lambda_{X}^2 e^{-\frac{M_X^2}{M^2}}= \int_{\Delta^2}^{s_0} ds
\left[\frac{1}{2}\rho_{11}(s)+\frac{\langle \bar{s}s\rangle^2}{6}
\rho_{22}(s)\right] e^{-\frac{s}{M^2}} \, .
\end{eqnarray}

 Differentiating  Eq.(8) with respect to  $\frac{1}{M^2}$, then eliminate the
 pole residue $\lambda_{X}$, we can obtain the sum rule for
 the mass,
 \begin{eqnarray}
 M_X^2= \frac{\int_{\Delta^2}^{s_0} ds
\frac{d}{d \left(-1/M^2\right)}\left[3\rho_{11}(s)+ \langle
\bar{s}s\rangle^2  \rho_{22}(s)\right]e^{-\frac{s}{M^2}}
}{\int_{\Delta^2}^{s_0} ds \left[3\rho_{11}(s)+ \langle
\bar{s}s\rangle^2  \rho_{22}(s)\right]e^{-\frac{s}{M^2}}}\, .
\end{eqnarray}

\section{Numerical results and discussions}
The input parameters are taken to be the standard values $\langle
\bar{q}q \rangle=-(0.24\pm 0.01\, \rm{GeV})^3$, $\langle \bar{s}s
\rangle=(0.8\pm 0.2)\langle \bar{q}q \rangle$, $\langle
\bar{s}g_s\sigma G s \rangle=m_0^2\langle \bar{s}s \rangle$,
$m_0^2=(0.8 \pm 0.2)\,\rm{GeV}^2$, $\langle \frac{\alpha_s
GG}{\pi}\rangle=(0.33\,\rm{GeV})^4 $, $m_s=(0.14\pm0.01)\,\rm{GeV}$
and $m_c=(1.35\pm0.10)\,\rm{GeV}$   at the energy scale  $\mu=1\,
\rm{GeV}$ \cite{SVZ79,Reinders85,Ioffe2005}.

 In the conventional QCD sum
rules \cite{SVZ79,Reinders85}, there are two criteria (pole
dominance and convergence of the operator product expansion) for
choosing  the Borel parameter $M^2$ and threshold parameter $s_0$.
In Refs.\cite{Wang0903,Wang0907}, we study the ${D}^\ast {\bar
{D}}^\ast$, ${D}_s^\ast {\bar {D}}_s^\ast$, ${B}^\ast {\bar
{B}}^\ast$ and ${B}_s^\ast {\bar {B}}_s^\ast$ molecular states in a
systematic way, the threshold parameters are
$s_0=(24\pm1)\,\rm{GeV}^2$, $(25\pm1)\,\rm{GeV}^2$,
$(138\pm2)\,\rm{GeV}^2$  and $(140\pm2)\,\rm{GeV}^2$ in the
$\bar{c}\gamma_\mu u \bar{d} \gamma^\mu c$,
   $\bar{c}\gamma_\mu s \bar{s} \gamma^\mu c$, $\bar{b}\gamma_\mu u \bar{d} \gamma^\mu
   b$   and $\bar{b}\gamma_\mu s \bar{s} \gamma^\mu b$ channels respectively;
   the Borel parameters are   $M^2=(2.6-3.0)\,\rm{GeV}^2$ and
   $(7.0-8.0)\,\rm{GeV}^2$ in the
hidden charm  and hidden bottom channels respectively. In those
regions, the two criteria of the QCD sum rules are satisfied. In
this article, we choose the interpolating current $J(x)$, which is a
special superposition of the scalar currents $\bar{c}(x)c(x)$ and
$\bar{c}(x)\gamma_\mu s(x) \bar{s}(x) \gamma^\mu c(x)$. So we can
take the same threshold parameter and Borel parameter as in the
channel $\bar{c}\gamma_\mu s \bar{s} \gamma^\mu c$, i.e.
$s_0=(25\pm1)\,\rm{GeV}^2$ and $M^2=(2.6-3.0)\,\rm{GeV}^2$.

The contributions from the different terms   in the operator product
expansion are shown in Fig.1, from the figure, we can see that the
dominant contribution comes  from the perturbative term and  the
operator product expansion is well convergent. In Fig.2, we show the
contribution from the pole term with variation of the Borel
parameter and the threshold parameter. The pole contribution is
larger than  $50\%$, the pole dominant condition is also satisfied.

Taking into account all uncertainties of the relevant  parameters,
finally we obtain the values of the mass and pole residue  of
 the narrow structure   $X(4350)$, which are  shown in Fig.3,
 \begin{eqnarray}
 M_X&=&(4.37\pm0.15)\,\rm{GeV} \, ,\nonumber\\
 \lambda_X&=&(4.1\pm0.8)\times 10^{-2}\,\rm{GeV}^5  \, .
 \end{eqnarray}
The prediction  is in good agreement with the experimental data
$M_X=(4350.6^{+4.6}_{-5.1}\pm 0.7)\,\rm{MeV}$ \cite{Belle4350}, the
$X(4350)$ may be a scalar $\bar{c}c-{D}_s^\ast {\bar {D}}_s^\ast$
mixing state. Other possibility, such as a scalar (tensor)
$c\bar{c}s\bar{s}$ tetraquark state is not excluded.

The nominal thresholds of the $D_s-\bar{D}_s$ and
$D^*_s-\bar{D}^*_s$
 are $M_{D_s\bar{D}_s}=3.937\,\rm{GeV}$ and
$M_{D_s^*\bar{D}^*_s}=4.225\,\rm{GeV}$ respectively \cite{PDG}, the
strong decays $X(4350) \to D_s\bar{D}_s  , D^*_s\bar{D}^*_s$ can
take place, so we can search for the $X(4350)$ in the $D_s\bar{D}_s$
and $D^*_s\bar{D}^*_s$ invariant mass distributions. The decay
channel $X(4350) \to   D^*_s\bar{D}^*_s$ has much smaller phase
space comparing with the  decay  channel $X(4350) \to D_s\bar{D}_s$,
the strong decay  $X(4350) \to D_s\bar{D}_s$ is of great importance.
By measuring the relative angular distributions of the pseudoscalar
mesons $D_s$ and $\bar{D}_s$, we can determine the spin of the
$X(4350)$.

 In Ref.\cite{4Coupling}, the light nonet
scalar mesons are taken as tetraquark states, and the strong
coupling constants among the light scalar mesons and pseudoscalar
mesons are calculated with the QCD sum rules. The numerical results
indicate that the values of the strong coupling constants for the
tetraquark states are always smaller  than the corresponding ones
for the $q\bar{q} $ states \cite{Colangelo03,Wang04}. In
Ref.\cite{Maiani20042}, Maiani et al take the diquarks as the basic
constituents, examine the rich spectrum of the diquark-antidiquark
states  with  the constituent diquark masses and the spin-spin
 interactions, and try to  accommodate some of the newly observed charmonium-like resonances not
 fitting a pure $c\bar{c}$ assignment. The predictions (also the Ref.\cite{Drenska0902}) depend  on  the assumption that the light
 scalar mesons $a_0(980)$ and $f_0(980)$ are tetraquark states,
 the  basic  parameters (constituent diquark masses) are
 estimated thereafter. If the  scenarios of the light nonet
scalar mesons  as the tetraquark states are robust, the  scalar
(tensor) $c\bar{c}s\bar{s}$ tetraquark state will have smaller
$D_s\bar{D}_s$ decay width than the corresponding ones of the
$c\bar{c}-D_s^*\bar{D}_s^*$ mixing state.

\begin{figure}
 \centering
  \includegraphics[totalheight=6cm,width=7cm]{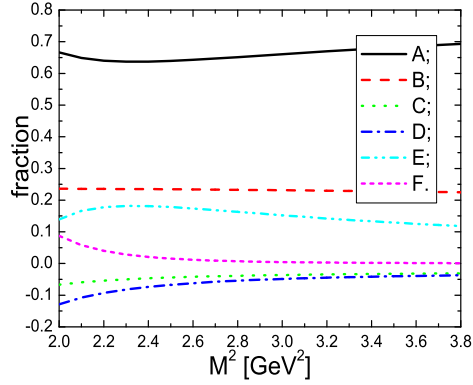}
     \caption{ The contributions from the different terms  with variation of the Borel parameter $M^2$ for $s_0=25\,\rm{GeV}^2$ in the
    operator product expansion. The $A$,
   $B$, $C$, $D$, $E$ and $F$ correspond to the contributions from
   the perturbative term,
$\langle \bar{s} s \rangle+\langle \bar{s}g_s\sigma G s \rangle$
term,  $\langle \frac{\alpha_s GG}{\pi} \rangle $ term, $\langle
\frac{\alpha_s GG}{\pi} \rangle $+$\langle \frac{\alpha_s GG}{\pi}
\rangle \left[\langle \bar{s} s \rangle +\langle \bar{s}g_s\sigma G
s \rangle+ \langle \bar{s}s \rangle^2\right]$ term, $\langle \bar{s}
s \rangle^2$+$\langle \bar{s} s \rangle\langle \bar{s}g_s\sigma G s
\rangle$ term and    $\langle \bar{s}g_s\sigma G s \rangle^2$ term,
respectively.      }
\end{figure}

\begin{figure}
 \centering
  \includegraphics[totalheight=6cm,width=7cm]{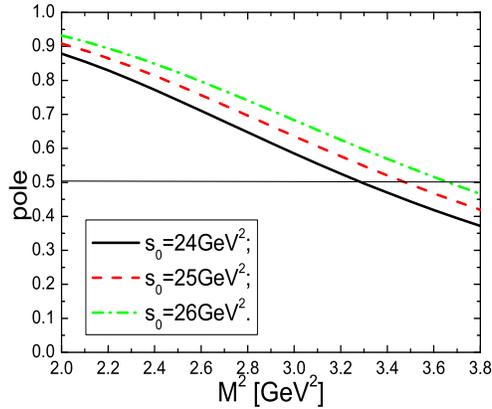}
    \caption{ The contribution of the pole term with variation of the Borel parameter $M^2$.  }
\end{figure}

\begin{figure}
\centering
\includegraphics[totalheight=6cm,width=7cm]{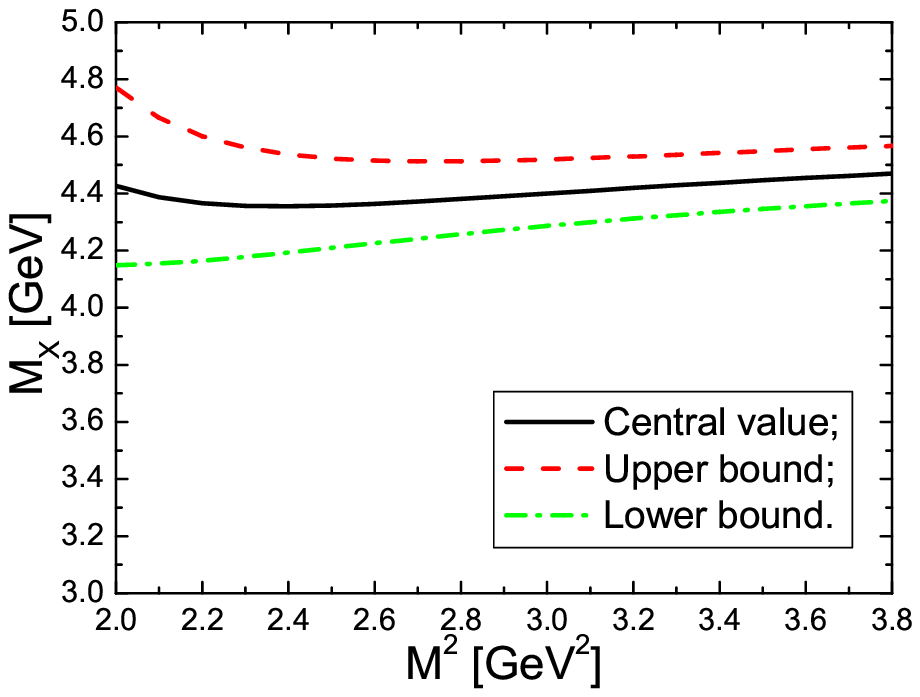}
\includegraphics[totalheight=6cm,width=7cm]{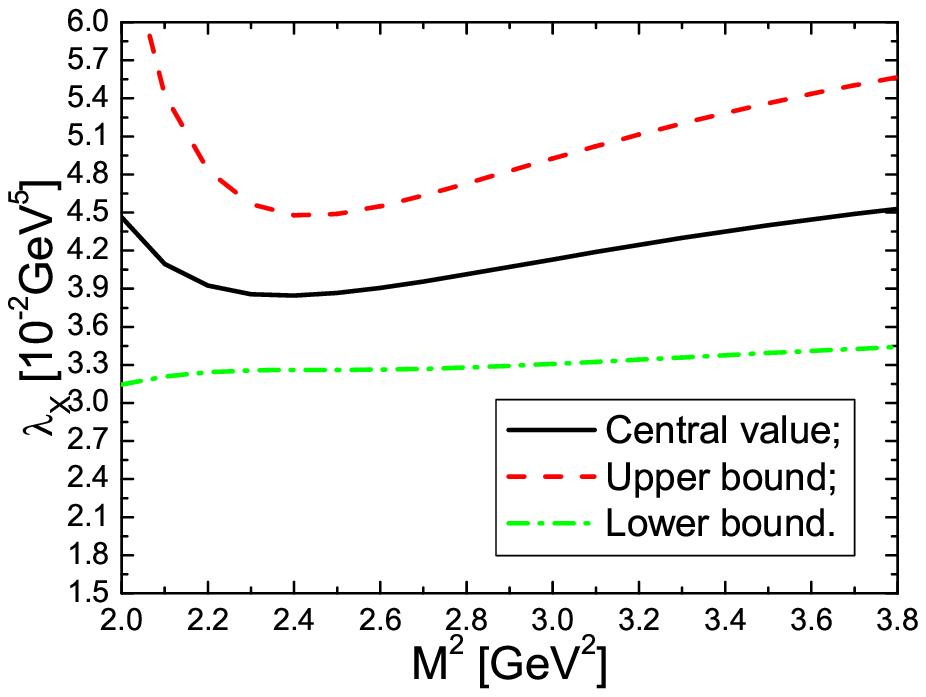}
  \caption{ The mass and pole residue of the $X(4350)$ with variation of the Borel parameter $M^2$.   }
\end{figure}

\section{Conclusion}
In this article, we assume that the $X(4350)$ is a scalar
$\bar{c}c-{D}_s^\ast {\bar {D}}_s^\ast$ mixing  state, and study its
mass using the QCD sum rules. Our prediction depends heavily on the
two criteria (pole dominance and convergence of the operator product
expansion) of the QCD sum rules. The numerical result is in good
agreement with the experimental data, the $X(4350)$ may be a scalar
$\bar{c}c-{D}_s^\ast {\bar {D}}_s^\ast$ mixing  state. Other
possibility, such as a scalar (tensor) $c\bar{c}s\bar{s}$ tetraquark
state is not excluded. We can search for $X(4350)$ in the
$D_s\bar{D}_s$ and $D^*_s\bar{D}^*_s$ invariant mass distributions,
especially the $D_s\bar{D}_s$. By measuring the relative angular
distributions of the pseudoscalar mesons $D_s$ and $\bar{D}_s$, we
can determine the spin of the $X(4350)$.

\section*{Acknowledgements}
This  work is supported by National Natural Science Foundation of
China, Grant Number 10775051, and Program for New Century Excellent
Talents in University, Grant Number NCET-07-0282, and the
Fundamental Research Funds for the Central Universities.

\end{document}